\documentclass[11pt,twoside]{article} 
\usepackage{asp2004}
\usepackage{epsf}
\usepackage{psfig}
\usepackage{lscape} 

\begin{document} 
\title{Modelling Alkali Line Absorption and Molecular Bands in Cool DAZs}
\author{Derek Homeier}
\affil{Institut f{\"u}r Astrophysik, Georg-August-Universit{\"a}t,
  Friedrich-Hund-Platz 1, D-37077 G{\"o}ttingen, Germany}
\author{Nicole Allard} 
\affil{Institut d'Astrophysique de Paris, CNRS, 
  98bis Boulevard Arago, 75014 Paris, France}
\author{France Allard} 
\affil{Centre de Recherche Astronomique de Lyon, 
  {\'E}cole Normale Sup{\'e}rieure, 69634 Lyon Cedex 07, France}
\author{Peter H.\ Hauschildt, Andreas Schweitzer}
\affil{Hamburger Sternwarte, Gojenbergsweg 112, D-21029 Hamburg,
  Germany}
\author{Phillip C.\ Stancil}
\affil{Department of Physics \& Astronomy, University of Georgia, 
  Athens, GA 30602-2451, USA}
\author{and Philippe F.\ Weck}
\affil{Department of Chemistry, University of Nevada, Las Vegas, 
  4505 Maryland Parkway, Box 454003, Las Vegas, NV 89154-4003, USA}

\begin{abstract} 
Two peculiar stars showing an apparent extremely broadened and strong
Na\,I\,D absorption have been discovered in surveys for cool white
dwarfs by Oppenheimer et al. (2001) and Harris et al. (SDSS, 2003). 
We discuss the nature of these objects using PHOENIX atmosphere 
models for metal-poor brown dwarfs/very low mass stars, 
and new white dwarf LTE and NLTE models for hydrogen- and
helium-dominated atmospheres with metals. These include complete
molecular formation in chemical equilibrium and a model for the alkali
resonance line broadening based on the damping profiles of Allard et
al. (2003), as well as new molecular line opacities for metal hydrides. 
First results of our calculations indicate good agreement with a 
hydrogen-dominated WD atmosphere with a Na abundance roughly consistent 
with a state of high accretion. We analyse deviations of the abundances 
of Na, K, Mg and Ca from the cosmic pattern and comment on implications
of these results for standard accretion scenarios.
\end{abstract}

\section{Introduction}
WD2356$-$209 is an apparent cool white dwarf identified in the
high proper motion-survey for halo white dwarf candidates of
\citet{OHDHS01}, with a spectrum showing
very unusual wide and deep absorption at 5000\,--\,6000\,{\AA}. Its
exotic properties have been confirmed by \citet{SalimHalo04}, who
found a combination of blue {\bv} and extremely red {\hbox{$V\!-\!I$}} 
colours
putting this star far outside of observed and theoretical sequences
of both hydrogen- and helium-atmosphere white dwarfs.  
Meanwhile, another star with a similar spectroscopic and photometric
appearance had been identified in Data Release 2 of the Sloan Digital
Sky Survey \citep{sdssWD03,sdssDR2WD}:
SDSS\,J\,13\,30\,01.13+64\,35\,23.8 (hereafter SDSS\,J1330+6435), which
is also showing high proper motion of 0\farcs193/yr. 
The prominent flux depression in $V$ in these stars has been suggested
to be due to a strong and heavily broadened Na\,I\,D doublet. 
Similar features have been observed in the coolest brown dwarfs of
spectral type T and identified with the resonance lines of the alkali
metals Na and K, broadened by van der Waals interaction with both
H$_2$ and He \citep{tsujiGl229,BurrowsMS00}. 
 These observations made clear that at line widths of more
than some 10\,{\AA} the standard impact theory of line broadening,
leading to a classical Lorentz profile of the far wings, could no
longer adequately reproduce the observed line shapes. 
More detailed quantum-mechanical calculations of the interaction
between alkali atoms and H$_2$ and He perturbers
\citep{BurrVolNaK,Alkalis03} have shown that the Lorentzian line
profile underestimates the absorption cross section at distances
beyond 2\,--\,3 FWHM from the line centre, out to
$\sim$\,2000\,{\AA}, while at very large detuning it significantly
overestimates it. 
In this paper we are exploring theoretical spectra for a
number of different stellar models, 
using our latest calculations of alkali line profiles, and evaluate
their capability to explain the observed features of WD2356$-$209 and
SDSS\,J1330+6435. 

\section{Model Atmospheres}
We have calculated models using the general stellar atmosphere code
\texttt{PHOENIX} v.~13.6 \citep{hbjcam99}. 
\texttt{PHOENIX} has been continuously developed to model the
atmospheres of the coolest stars, brown dwarfs and gas giant planets, 
incorporating an equation of state (EOS) that treats over 600 molecular
species in chemical equilibrium (CE) and includes the formation of liquid
and solid condensates \citep{LimDust}. 
We are using the unified theory of \citet{Alkalis03} to predict the
complete spectrum of the resonance lines of the alkali metals Li, Na,
K, Rb and Cs from the centre to the extreme wings, however with normal
abundance patterns only Na and K are found to have widths beyond some
tens of {\AA}, 
and only for these perturbations due to both neutral He and H$_2$ in
two interaction geometries are available at this time. 
These models have been successfully applied to the highly transparent
and dense atmospheres of cool T dwarfs, the only astrophysical
objects in which such lines have been observed prior to the two stars
that are the topic of this paper \citep{Alkalis03,KISat04}. 
The colours of WD2356$-$209 and the blue spectrum of SDSS\,J1330+6435
suggest that the $\lambda$\,4227\,{\AA} Ca\,I line is also an
important absorber, but for this earth alkali metal no detailed
calculations of the line spectrum are available yet. We have tried to
simulate the broadening of this line by using the Na\,I data, but
multiplying the damping constant with 60$^{0.4}$ to account for its
non-hydrogenic configuration. 

In other respects though, T dwarf spectra disagree very strongly with
the observed signatures of WD2356$-$209 and SDSS\,J1330+6435: they show
equally prominent absorption from the K\,I
$\lambda\lambda$\,7667,7701\,{\AA} resonance doublet as from Na, 
and they have extremely red optical-to-near IR
spectra, with the optical flux almost completely quenched bluewards of
the Na\,I\,D doublet. Objects this cool could therefore not account
for the quite blue {\bv} colours of WD2356$-$209. 
For any brown dwarf or very low mass-star of higher temperature
($T_\mathrm{eff} \ga 1400$\,K) however the strength of alkali line
absorption rapidly decreases, as the atmospheres become more opaque
due to strong absorption from molecular bands and condensates. 
Keeping a strong signature of the resonance lines in atmospheres of
higher temperature would require a metallicity significantly below
the solar value. The thusly reduced amount of molecules and dust and
would increase the transparency of the atmosphere, allowing the atomic
lines to form again under conditions of high density and pressure.  
To explore this possibility we have therefore calculated a grid of
dwarf atmosphere models at $T_\mathrm{eff}$ up to 3000\,K and
[Fe/H] down to $-$\,5. For the main alternative explanation, these
stars indeed being white dwarfs with a high abundance of photospheric
metals, we computed models with log\,$g$ of 7.0\,--\,8.5 and hydrogen-
or helium-dominated atmospheres. 
For all models the partial pressures of all molecules in
the EOS of \citet{LimDust} have been calculated in full CE, although
in the hydrogen-rich WD models chemistry is dominated by only a few
hydrogen compounds, and in helium-rich ones of course not many
molecules form at all. The models do not include, on the other hand,
the effects of a non-ideal EOS that may become important in white
dwarfs this cool and dense \citetext{cf.\ also Kowalski, Mazevet \&\
  Saumon, this vol.!}.  

\section{Comparison with Models}
\subsection{The Nature of the Dwarf(s)}
\begin{figure}[htbp]
  \centering
  \plottwo{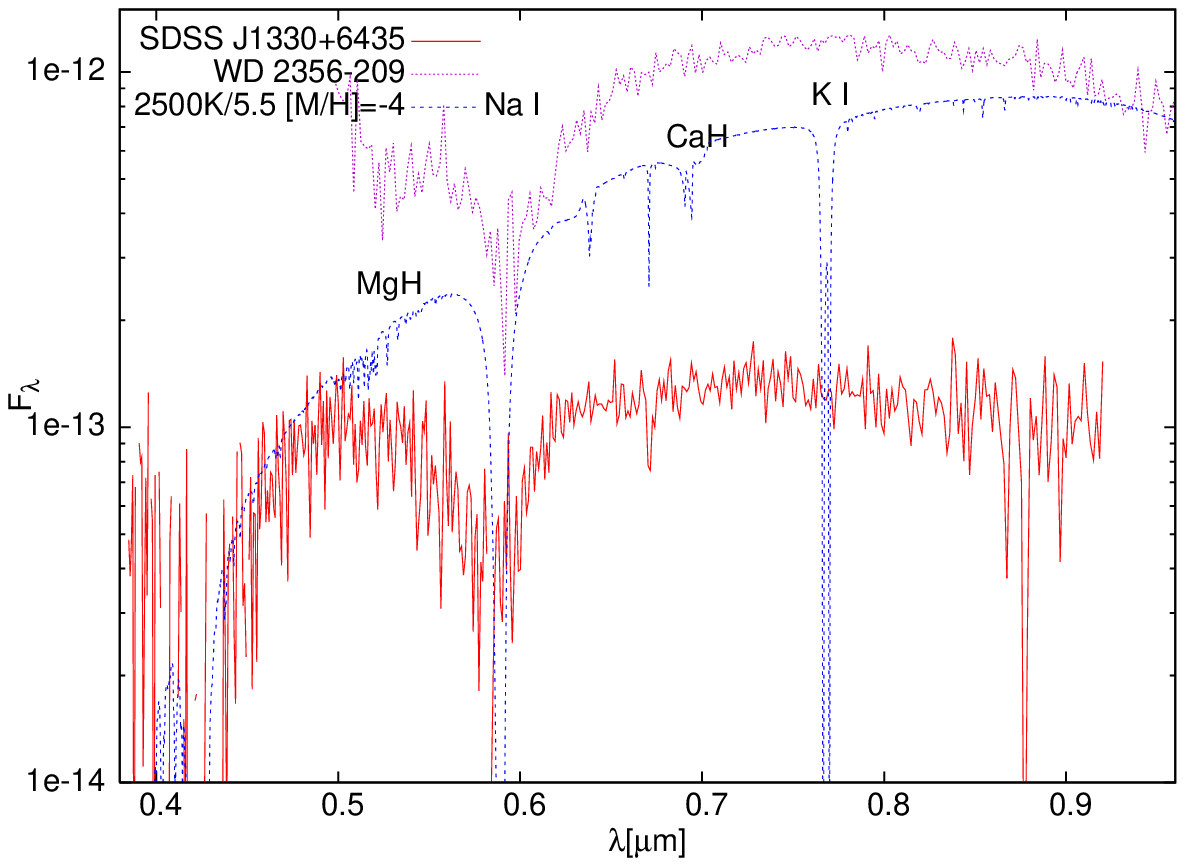}{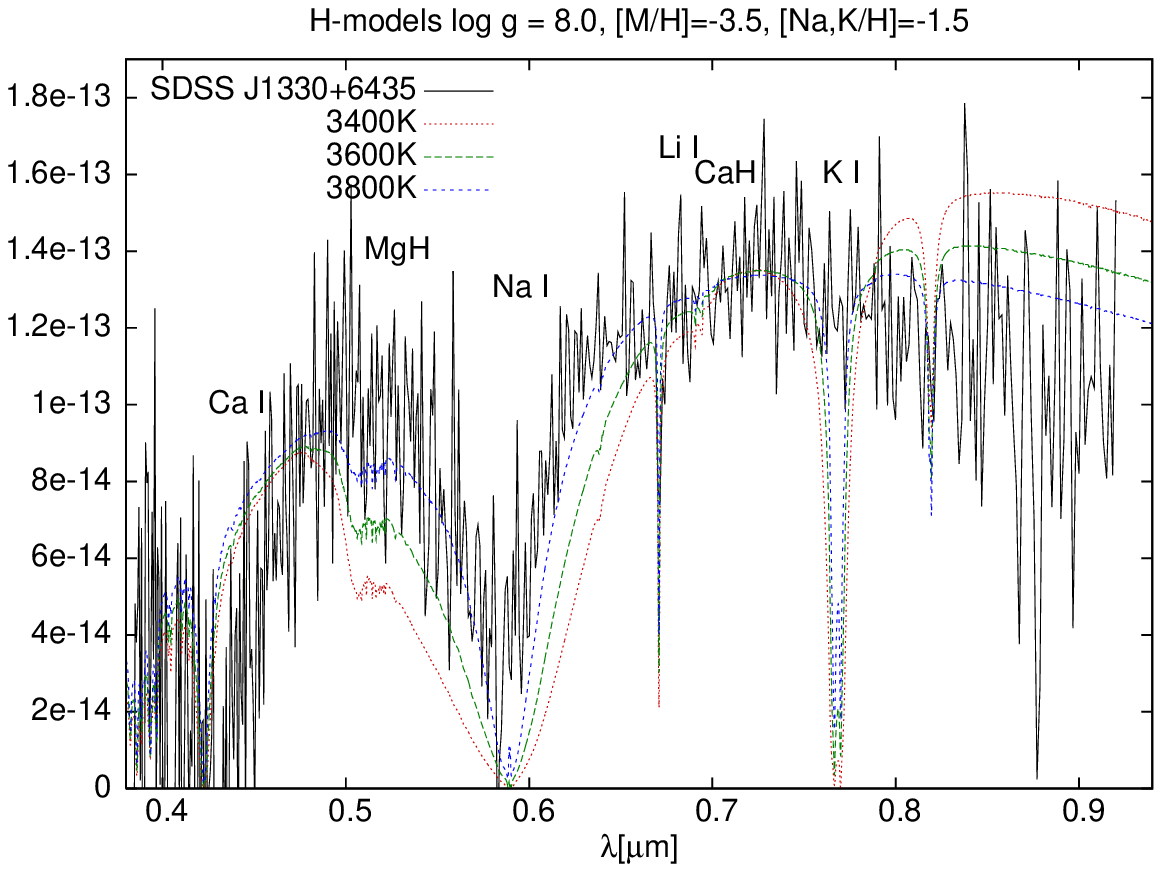}
  \caption{a:  (left) Comparison of a [Fe/H]\,=\,-$-$4 VLM-star model
    to the spectra of WD2356$-$209 and SDSS\,J1330+6435. The SDSS
    spectrum has been averaged with a 4-pixel boxcar filter to
    $R\approx 1000$. 
    b: (right) Fit of hydrogen-rich WD-models to the spectrum of
    SDSS\,J1330+6435. Principal absorption features are indicated}
  \label{fig:esdLmodel}
\end{figure}

Figure \ref{fig:esdLmodel}a shows a representative spectrum for an
extremely metal-poor model in the L-dwarf temperature range, as might
be conceived for a Population\,III star just above the
hydrogen-burning limit. It is immediately apparent that, without the
dust opacity dominating normal L-dwarf spectra, the alkali lines can
be strong enough to qualitatively explain the observed absorption
features. However the line profiles fail to reproduce the observed
shapes in detail, and show still too red optical$-$near-IR colours. We
conclude that a low mass-star or brown dwarf-type of model can be
excluded for these two stars. 
This makes the most likely classification of these objects white
dwarfs with traces of metal in an either helium- (DZ) or
hydrogen-dominated atmosphere (we have denoted the latter as DAZ, in
analogy to other hydrogen-rich white dwarfs showing metal absorption
lines besides the Balmer series of hydrogen, although due to the
extremely low temperatures in this case, Balmer lines would probably
not be observable even at the highest resolutions). 

\subsection{White Dwarf Models}
\subsubsection{Hydrogen-dominated atmospheres}
A series of fits to WD2356$-$209 to hydrogen-LTE
atmospheres is presented in Fig.~\ref{fig:esdLmodel}b. The strong
dependence of the alkali lines on temperature, becoming quickly weaker
with increasing $T_\mathrm{eff}$, is evident. 
Test models for solar abundance patterns predicted the formation of
stronger molecular bands, chiefly due to MgH and CaH, than are present
in the observations. Although the spectrum of WD2356$-$209 shows an
apparent additional opacity source in the blue wing of the Na\,I
doublet, its position does not quite agree with the modelled band,
while SDSS\,J1330+6435 only shows a marginally detectable feature. 
In addition, CaH bands should be visible around 7000\,{\AA} if the
elemental Ca abundance were in the same ratio to Na as in the Sun. The
models shown here therefore have Na and K abundances set to $-$1.5\,dex
relative to solar, and all other metals to $-$3.5. 
As can be seen, even at $T_\mathrm{eff} < 4000$\,K quite high Na
abundances are required to reproduce the strength of the D lines. 
In contrast the K\,I doublet, which is marginally if at all detectable
in the observations, is predicted to be much stronger at that
abundance. [K/H] therefore seems to have a lower abundance more
similar to the other metals. 

\subsubsection{Helium-rich models}
\begin{figure}[htbp]
  \centering
  \plottwo{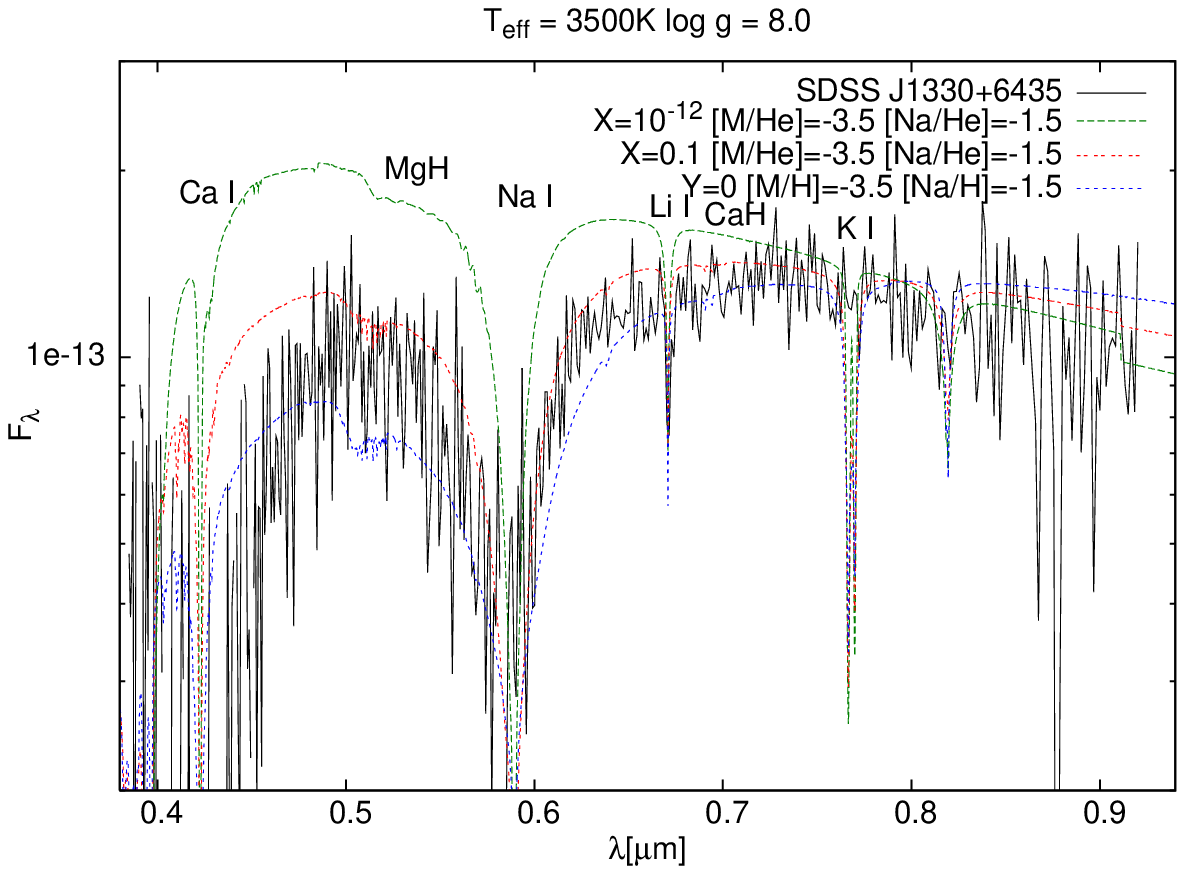}{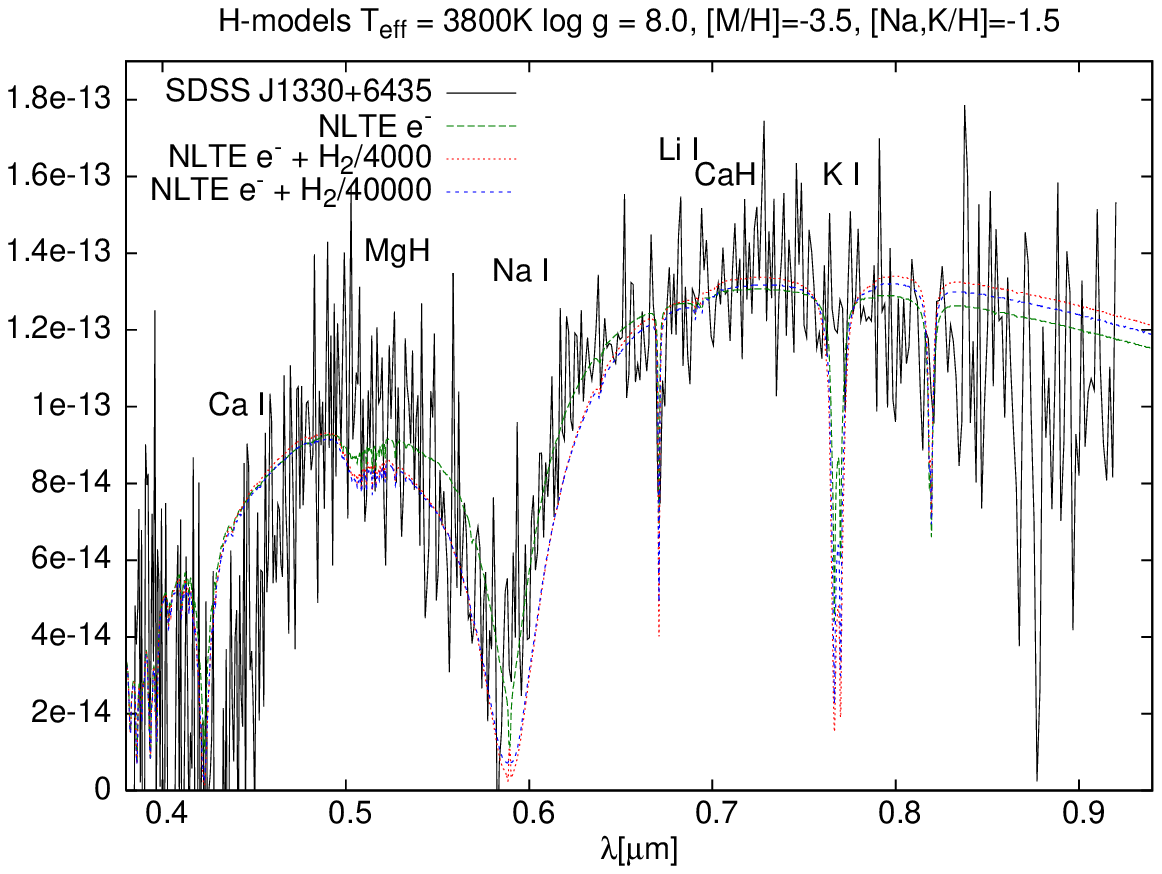}
  \caption{a:  (left) Comparison of models with different H/He number
    ratios for SDSS\,J1330+6435. 
    b: (right) NLTE effects on H-atmosphere models. High departure
    ratios in the upper atmosphere may be responsible for narrower
    line cores (long-dashed plot), however collisional deexcitation
    due to H$_2$ would most likely restore LTE (cf.\ text)}
  \label{fig:WDmodels2}
\end{figure}

In addition to the degeneracy of solutions with respect to
$T_\mathrm{eff}$ and metal abundances, the main constituent in the
atmosphere is also unknown. 
The effect of a varying H/He mixture is illustrated in
Fig.~\ref{fig:WDmodels2}a.  We find a slight improvement of the
line fit with moderate amounts of helium, e.\,g.\ 50\,--\,90\,\%. 
Infrared flux is also more suppressed, due to the stronger H$_2$\,-\,He
collision induced absorption (CIA), with increasing helium
abundance even up to H/He\,=\,10$^{-12}$. 
These models thus show rather blue synthetic IR colours of 
{$V\!-\!H \la 2.0$}, as compared 3.0\,--\,3.5 for pure hydrogen
models and an observed {$V\!-\!H = 2.75$} for WD2356$-$209 
\citetext{Bergeron et al., this vol.}. 
The best fits seem to be obtained with H/He of order unity. 
We did not run any completely hydrogen-free models, which should
eventually remove the CIA beyond 1\,$\mu$m. {Bergeron
  et al.} report a good fit with such models, although they only use
standard Lorentzian broadening for the Na\,I doublet. 

\subsubsection{NLTE Effects}
The line profiles particularly in our pure-hydrogen LTE models appear 
generally too deep in the near wings out to a few 100\,{\AA},
indicating that either the far wing is too weak compared to the core
and near wing in our line profiles, or for some reason Na\,I
absorption is overestimated in the upper atmosphere, where the inner 
parts of the lines form.  NLTE
effects could contribute to the latter. 
Departures from LTE are typically observed in white dwarfs of
$T_\mathrm{eff}\ga 20\,000$\,K, and might seem unlikely in dwarfs this
cool and dim. However, the highly neutral atmosphere of ultracool
dwarfs leaves almost no free electrons as the main source of
collisional deexcitation and could thus favour NLTE effects. 
Alkali line 
absorption under such circumstances has been studied by
\citet{nlteNaI} for the transit planet
HD\,209458b. 
Due to the extreme paucity of free electrons the dominant species,
molecular hydrogen, will most likely become the driving force to restore 
LTE. This effect can not be precisely modelled, since no
cross-sections for H$_2$\,--\,Na collisions are 
available yet. \citeauthor{nlteNaI} have discussed 
two limiting cases, including only $e^-$ collisions, and including
H$_2$ collisions with the same cross-sections as for
$e^-$. Theory suggests a much smaller H$_2$
cross-section, scaling with the inverse
mass ratio. We have therefore
also calculated NLTE models using 4000 and 40\,000
times smaller cross-sections for H$_2$. 
The test models in Fig.~\ref{fig:WDmodels2}b with $e^-$
only show very significant deviations from the LTE line shape, 
but even with the lowest estimate for H$_2$
collisional rates LTE is almost completely restored. 
NLTE effects are thus probably not the cause of the
discrepancies between models and observation. 

\subsubsection{Limitations of the models}
Since LTE does seem to hold within the line-forming regions, actual
depletion of elemental Na by gravitative diffusion could be another
explanation for the missing opacity in the deep line cores. 
This is not an uncommon effect in white dwarf atmospheres, but a
detailed calculation of it is beyond the scope of our 
current models. We also have to caution, however, that the
calculations for the line spectra we used were designed to 
be used with perturber densities of up to
10$^{19}$\,cm$^{-3}$, whereas at the higher densities of 
white dwarf atmospheres the single-perturber approximation 
breaks down \citep{Alkalis03}, 
leading to an underestimation of the far wing part of the line. 

\section{Conclusions} 
With the low resolution and $S/N$ of both observations and the
uncertainties of our current models it is not possible to draw strong
conclusions about the atmospheric conditions in these white
dwarfs. E.\,g.\ the presence of hydride bands cannot be clearly
established, which would allow a better determination of elemental
abundance ratios, and confirm whether hydrogen is an important
constituent of the atmosphere. The quite high Sodium abundance
required in the DAZ scenario, which might be beyond what interstellar
accretion could provide, is an obvious argument against such
models. The apparent overabundance of Na relative to Ca, and perhaps
also Mg and K, also disagrees with selective dust-accretion models,
which predict rather lower abundances of the volatile Alkalis
\citep{2003ApJ...596..477Z}. However, DZ models 
would face the same irregular abundance patterns, except
that hydride bands are not expected. Also the problem of avoiding the
accretion of hydrogen from the ISM is still unresolved. 
With the current knowledge, a model with He/H ratio of order unity
might best reconcile both observed features and the problems with the
extremely short diffusion timescales of pure hydrogen atmospheres. 

Observations that could resolve molecular bands, and
possibly lines from higher excitation and ionisation levels for 
better constraints on temperature, would be of great help in
deciding between these different interpretations. 
New theoretical work on the behaviour
of atomic lines at very high perturber densities should also both
benefit from and aid in the study of these unusual white dwarfs.

\acknowledgements{We we are very grateful to Ben Oppenheimer and Hugh Harris for
  providing access to electronic versions of their spectra, 
  and to Travis Barman for discussing the effect of collisional
  rates. 
  D.~H.\ thanks Holger Boll for kind hospitality during the workshop.
  Support for D.~H.\ from the National Science
  Foundation of the USA under NFS grant N-Stars RR185-258 to the
  University of Georgia is gratefully acknowledged. 
  Models presented in this work are based in part on calculations
  performed at the NERSC IBM SP at LBNL with support from the US 
  Department of Energy.}



\end{document}